\documentclass[twocolumn]{aastex63}

\usepackage[utf8]{inputenc}
\usepackage{natbib}
\usepackage{graphicx}
\usepackage{xcolor}
\pagecolor{white}
\usepackage{amsmath}
\usepackage{amssymb}
\usepackage{makecell}

\begin{document}
\title{On the Existence of Fast Modes in Compressible Magnetohydrodynamic 
Turbulence}

\author{Zhaoming Gan}
\affiliation{New Mexico Consortium, Los Alamos, NM 87544, USA}
\affiliation{Los Alamos National Laboratory, Los Alamos, NM 87545, USA}

\author{Hui Li}
\affiliation{Los Alamos National Laboratory, Los Alamos, NM 87545, USA}

\author{Xiangrong Fu}
\affiliation{New Mexico Consortium, Los Alamos, NM 87544, USA}
\affiliation{Los Alamos National Laboratory, Los Alamos, NM 87545, USA}

\author{Senbei Du}
\affiliation{Los Alamos National Laboratory, Los Alamos, NM 87545, USA}

\begin{abstract}
We study the existence and property of Fast magnetosonic 
modes in 3D compressible MHD turbulence by carrying out 
a number of  simulations with
compressible and incompressible driving conditions.
We use two approaches to 
determine the presence of Fast modes: mode decomposition based on spatial variations 
only and spatio-temporal 4D-FFT analysis of all fluctuations. 
The latter method enables us to quantify fluctuations that satisfy the dispersion
relation of Fast modes with finite frequency. 
Overall, we find that the fraction of Fast modes identified via spatio-temporal 
4D FFT approach in total fluctuation power 
is either tiny with nearly incompressible driving or $\sim 2\%$
with highly compressible driving. 
We discuss the implications of our results for understanding the compressible 
fluctuations in space and astrophysics plasmas. \\
\end{abstract}

\section{Introduction}

Magnetized plasma systems, such as magnetic fusion experiments  
\citep[e.g.,][]{Diamond_2005}, 
solar wind \citep[e.g.,][]{Matthaeus_1982}, interstellar medium \citep[e.g.,][]{Armstrong_1995}, 
and intracluster medium \citep[e.g.,][]{Hitomi_2018}, 
are often turbulent. 
Magnetohydrodynamics (MHD) is generally  employed to describe turbulence 
in such systems. 
Understanding the nature of compressible fluctuations 
is also
critical in describing the relationship between the weak and strong turbulence limits
\citep[e.g.,][]{Chandran_2005, Galtier_2009, Meyrand_2016, Galtier2018}, 
because the cascade process, anisotropy, and energy spectrum are likely all modified by
compressible effects. 

The eigenmodes in a compressible MHD system are  
different from those in the incompressible limit. We choose to focus on
Fast magnetosonic modes or Fast modes in this study. 
Fast modes are often included in global solar wind models, 
critical in preferential ion heating at kinetic scales 
through turbulent cascade \citep[e.g.,][]{cranmer_proton_2012}.
They may also play an important role in interpreting 
the recent \textit{Voyager 1} observations 
\citep{Zank_2017,Zank_2019, Burlaga_2018, Zhao_2020}.
Fast modes have also been used in
accelerating particles in solar flares
\citep[e.g.,][]{Miller_1996}, cosmic rays \citep[e.g.,][]{schlickeiser_cosmic_2002}, and in 
scattering cosmic rays in the interstellar medium \citep[e.g.,][]{Yan_2002}. 
Furthermore, it is postulated that Fast mode turbulence
is effective in producing stochastic particle acceleration in 
various high-energy astrophysical systems 
as well  \citep[e.g.,][]{Dermer_1996, Li_1997, Demide_2020}. 

Earlier studies have examined the fraction of compressible modes in MHD turbulence
and their cascade properties. The primary method 
is to use only the {\em spatial variations}
of various variables and decompose them into three eigenmodes -- Fast, Slow and Alfv\'en modes
(see e.g. \citealt{marsch_acceleration_1986, cho_compressible_2003, zhang_occurrence_2015, 
yang_coexistence_2018}). 
The effects of turbulence driving conditions
have been examined in \citet{makwana_properties_2020}, and
they found that the Fourier power in Fast modes could be up to $\sim$30\% of 
the total turbulent power if the turbulence driving is highly {\it compressible}, 
and the power in slow modes is even more significant. 

Another interesting approach in identifying  Fast modes in a turbulence
simulation was discussed in \cite{Svidzinski_2009}.  They demonstrated that
a spatio-temporal analysis method can be used to decompose the fluctuations
(say magnetic fields) using Fast Fourier Transform (FFT) 
in both spatial and temporal domains, if a large 
number of data volumes at different time slices are stored to resolve both low
and high frequencies (we will refer this approach as 4D FFT). 
Similar approaches have also been applied to MHD turbulence
simulations 
\citep[e.g.,][]{dmitruk_low-frequency_2007,dmitruk_waves_2009, clark_di_leoni_spatiotemporal_2015, 
meyrand_direct_2016, andres_interplay_2017, lugones_spatio-temporal_2019, 
Yang_2019, Brodiano_2021}, 
as well as in hybrid kinetic simulations \citep{Markovskii_2020}.
{ \cite{dmitruk_low-frequency_2007,dmitruk_waves_2009} examined spatio-temporal turbulence signals in the incompressible regime.}
The existence of compressible modes (Fast and Slow waves)
that satisfy dispersion relations in the frequency-wavenumber ($\omega$ vs ${\bf k}$) 
space \citep[e.g.,][]{andres_interplay_2017, Yang_2019}
has { also} been demonstrated. 
{ Following these earlier studies, we build similar 4D FFT routines to analyze MHD turbulence.}
In particular, 
the recent paper by \cite{Brodiano_2021} is most similar to the study
presented here. They studied the effects of how the compressible vs. 
incompressible driving affects the presence of waves in turbulence. 
They concluded that the system is mainly dominated by the nonlinear
fluctuations (i.e., not waves).

In this paper, we aim at addressing the detailed properties of
Fast modes in compressible MHD turbulence, particularly the existence 
of finite frequency waves and how they vary under different turbulence driving
conditions.
In \S\ref{sec:model} we describe 
our simulation set-ups and various runs we performed. 
In \S\ref{sec:results} we present our analysis and results of numerical simulations. 
Conclusions and implications of our results are given in \S\ref{sec:discussion}.

\section{Model Description} \label{sec:model}
\subsection{Numerical Schemes}

To address the existence and properties of Fast modes in compressible MHD turbulence,
we solve the time-dependent ideal MHD equations numerically in a 
three-dimensional Cartesian coordinate system ($x,y,z$) 
using the code \texttt{ATHENA++} 
\citep{Stone_2020}.
We introduce ${\bf f_{\rm v}}$ and ${\bf f_{\rm B}}$ terms
in the momentum and induction equations 
as turbulence driving forces on 
velocity and magnetic fields, respectively. 
Both ${\bf f_{\rm v}}$ and ${\bf f_{\rm B}}$ take the form of
${\bf A} \sin({\bf k}\cdot{\bf x} + \phi)$ across the computational domain 
with wave numbers ${\bf k}$ that satisfy periodical boundary,
and randomly selected phases $\phi$, 
and amplitudes ${\bf A_{v}}$, ${\bf A_{B}}$ for velocity and magnetic fields, respectively. 
The amplitude is decomposed into components 
parallel and perpendicular to the background magnetic field 
${\bf B}_0$ as ${\bf A_\parallel}$ and ${\bf A_\perp}$.
For velocity driving, we further define a free parameter $f_c$ as 
$|{\bf A_\parallel}| = |{\bf A_\perp}|\cdot f_c/(1-f_c)$ so that
$f_c = 0, 1$ represent the incompressible and fully compressible driving
limits, respectively. 
All runs have uniform background density and magnetic fields in 
which we choose our 
normalization as $\rho_0 = 1$ and ${\bf B}_0 = [8, 0, 0]$,
so the characteristic Alfv\'en speed is 
$v_{\rm A} = 8$. 
We use an isothermal equation of state.
The initial mean velocity is set to $0$, and for all runs, we set a uniform initial sound speed $c_s=\sqrt{p/\rho}$
using $\beta= 8 \pi p/B^2 = 0.4$. 
We apply random driving that follows the Ornstein-Uhlenbeck process 
with a correlation time $t_{cor} = 0.5 \tau_A$ \citep{eswaran_examination_1988}\textcolor{black}{, 
where $\tau_A = L_0/v_A$} with $L_0$ being the box size along ${\bf B}_0$.  
All driving and/or injection occurs at specific wavenumbers that 
have $0 < |{\bf k}_{\rm inj}| \leq 2$.
The range of turbulent Mach number $M_{turb} \equiv \delta v/c_s$ is between  0.12 and 0.18 for the simulations discussed here.

\begin{figure*}
\centering
\includegraphics[width=0.80\textwidth]{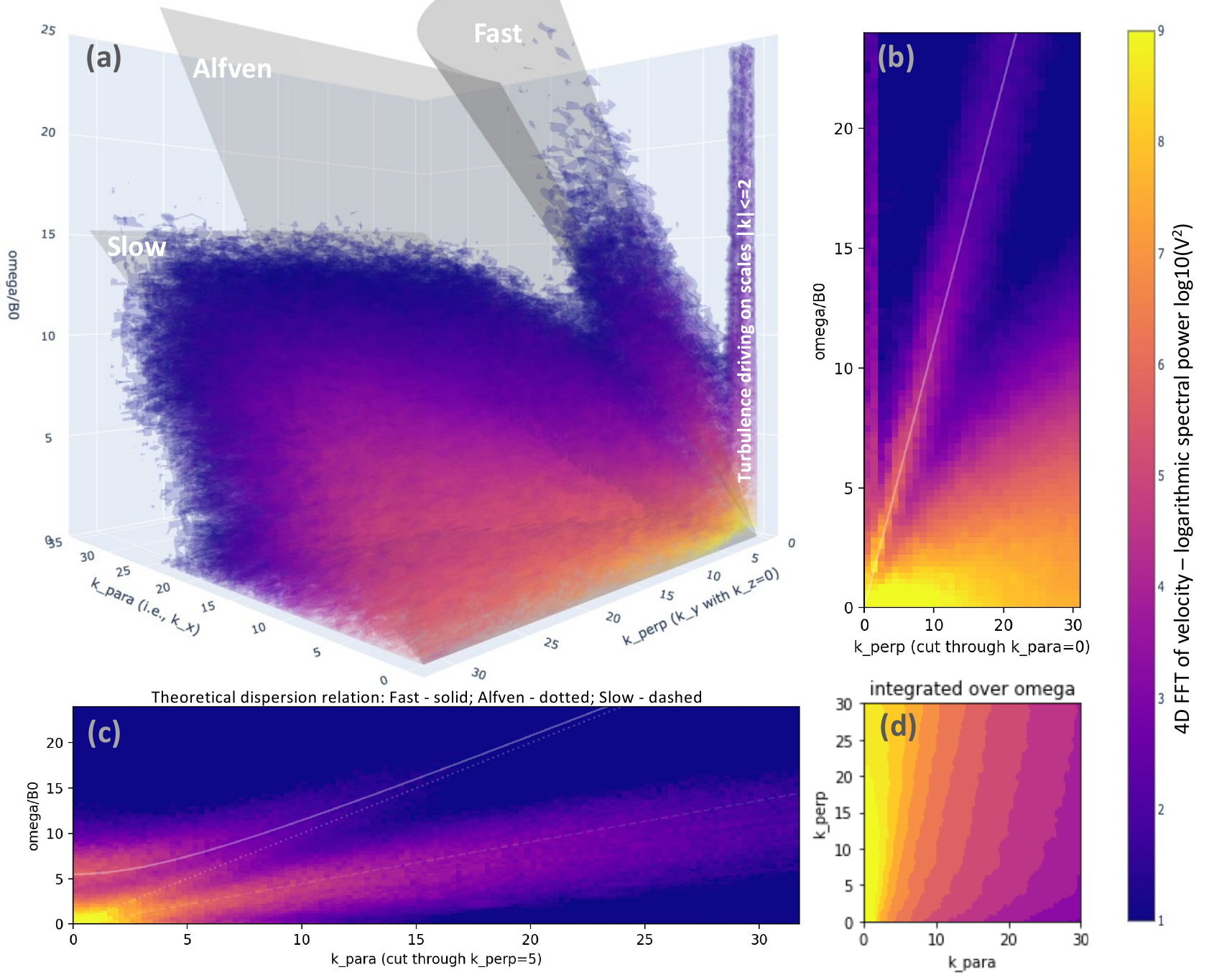}
\caption{Spatio-temporal (4D FFT) spectrum of run B
in the $\omega$ vs. $k_\perp$ and $k_\parallel$ domain, showing the  
Fast, Alfv\'en, Slow components that are indicated by the
gray surfaces shown in Panel (a), as well as fluctuations we call 
``non-waves". 2D cuts are shown in Panels (b) and (c) to 
emphasize the three wave branches and the non-wave component. 
Panel (d) depicts the power distribution in the $k_{\parallel}-k_{\perp}$ plane 
by integrating the spectrum over all frequencies. 
}
\label{fig1}
\end{figure*}

\subsection{Mode decomposition Methods}  \label{sec:mode_decomposition}

We describe two methods to identify various wave modes in turbulence. 
The first one is the ``spatial-only" method described in \cite{cho_compressible_2003,
yang_coexistence_2018}, which uses the velocity output at each time frame
to identify the fraction of Alfv\'en, Fast and Slow modes, according to the polarization features of velocity 
predicted by linear wave theory.
This approach 
relies on the fact that any spatial variations
at any given timeframe
can be projected onto three eigenvectors {\it mathematically}. 
The physical interpretation of this approach, however, is a bit 
unclear. It certainly makes sense physically when the
fluctuation amplitudes are small and nonlinear interactions
among different modes are not dominant.
But for strong MHD turbulence, the nonlinear interactions become
important. It is unclear this approach alone can be used to 
decide the fractions of various wave components. 

The other method we used is the spatio-temporal 
Fourier transform (4D FFT)
which allows us to obtain both the spatial and frequency information of variations. 
By storing the data outputs frequently with sufficient 
time resolution ($\sim 100$ frames within each $\tau_A$ up to 
$\sim 10~\tau_A$), 
we are able to resolve the dispersion relation of linear waves. 
This ensures that both the high and low  frequencies are captured 
(they are also related to the 
spatial resolution and box size). 
To reduce the data volume, we split the 4D 
Fourier transform into two steps: First, we make 3D spatial 
FFT for a variable (e.g., velocity), and 
store the intermediate data in a time series. Second, we perform 
the fourth (temporal) layer of the spatio-temporal Fourier transform 
(for individual frequencies) by integrating the intermediate data over time, 
and save the results in a series of individual frequencies for further analysis. The outcome of these analyses will be spectral
power populated within the 4D $\omega - {\bf k}$ volume. 
This approach 
has the advantage
of assessing whether the Fast, Slow, Alfv\'enic fluctuations indeed 
satisfy their respective dispersion relations.

The three eigenmodes delineate three isosurfaces in the 
$\omega - {\bf k}$ volume. To determine whether 
certain spectral power belongs to a specific eigenmode,
we allow $\pm 10\%$ 
deviation in $\omega$ for a specific ${\bf k}$ 
to compensate for two effects. First,
there is a well-known issue of spectral leakage 
in Fourier transform when 
periodic signals are truncated unevenly \citep{harris_use_1978}, 
which is 
the case in the 
temporal dimension. (We employ periodic boundaries in the 
spatial dimensions.)  To suppress the spectral leakage, the Hanning 
window function is used in the temporal dimension. 
Second, there is the possible frequency broadening due to 
nonlinear interactions. 
We have also tried to use other ``width" 
$\pm 3\%$ and $\pm 40\%$, but 
the main conclusions do not change
qualitatively.

\section{Results} \label{sec:results}

We present results from three \textcolor{black}{runs A, B, and C} which 
differ in their turbulence driving:
both A and C have only velocity driving (${\bf f_{\rm v}}$) but A has incompressible ($f_c = 0$) 
while C has highly compressible driving ($f_c = 0.9$). 
Run B has $f_c = 0$, but with both
${\bf f_{\rm v}}$ and ${\bf f_{\rm B}}$. 
All simulations use $512^3$ resolution. 
Both A and C have a cubic simulation box of size $2\pi\times 2\pi \times 2\pi$,
and B has an elongated box of size $8\pi\times2\pi\times2\pi$. 
All analyses are performed after simulations have
reached quasi-steady state, typically after several $\tau_{\rm A}$.

First, we show the outcome of 4D FFT analyses in identifying
various waves as depicted in Figure \ref{fig1}. Panel (a) shows
the spatio-temporal spectrum of 
run B with an elongated box in a 3D representation of
$\omega$ vs. $k_x$
and $k_y$ with  $k_z = 0$ for this plot.
We note that $k_{\parallel} \simeq k_x$ and 
$k_\perp \simeq (k_y^2 + k_z^2)^{1/2}$.
It is clear that
fluctuations have cascaded to higher $k$ and with finite frequency.
Several main features can be identified. 
The theoretical dispersion planes are marked as gray surfaces 
for Fast, Alfv\'en and Slow modes. The vertical feature along $\omega$ axis 
at small $|{\bf k}| \leq 2$  is due to driving. 
Panel (b) shows a 2D cut with $k_x=0$ in which Fast mode is 
most easily identified, given their finite frequencies. 
In this limit, both Alfv\'en and Slow modes have zero frequency. 
Panel (c) shows another 2D cut with $k_\perp = 5$.
This brings out
all three wave branches (as marked by the three white lines).  
Panel (d) shows the power distribution in the $k_\parallel-k_\perp$ 
plane by integrating the spectrum over all frequencies. 
This distribution is similar to the previous studies that demonstrate
the anisotropic cascade in $k$-space \citep[e.g.,][]{Chhiber_2020}. 
Note that, in Panels (b) (c) (d), we have integrated all possible 
combination of $k_y$ and $k_z$ for a given $k_\perp$ 
to capture all the spectral power. 

It is interesting to see that there is limited amount of power (to be quantified 
later) in fluctuations that cascade both to higher $k$ and higher 
$\omega$ and satisfy the dispersion relations for the three eigenmode
branches, as most easily seen in Panels (c) and (b). 
Slow modes tend to cascade further in $k$ than both the Alfv\'en 
and Fast modes, as corroborated by Panel (a). 

The most prominent feature is the large fraction of fluctuation power
{\it not} on any of the three wave branches, which is clearly shown by the lower right
region in Panel (b) and the regions in-between three wave surfaces in Panel (a). 
These fluctuations contain finite ${\bf k}$ and finite $\omega$ even
though the highest concentration 
of power tends to be in the low $\omega$ region.  
We call them the ``non-wave" component, which 
judging from these plots,
contains the most amount of power. By inference, most of the power
contained in the $k_\perp$ cascade shown in Panel (d) resides in this ``non-wave" 
component as well.

\begin{figure}
\centering
\includegraphics[width=0.475\textwidth]{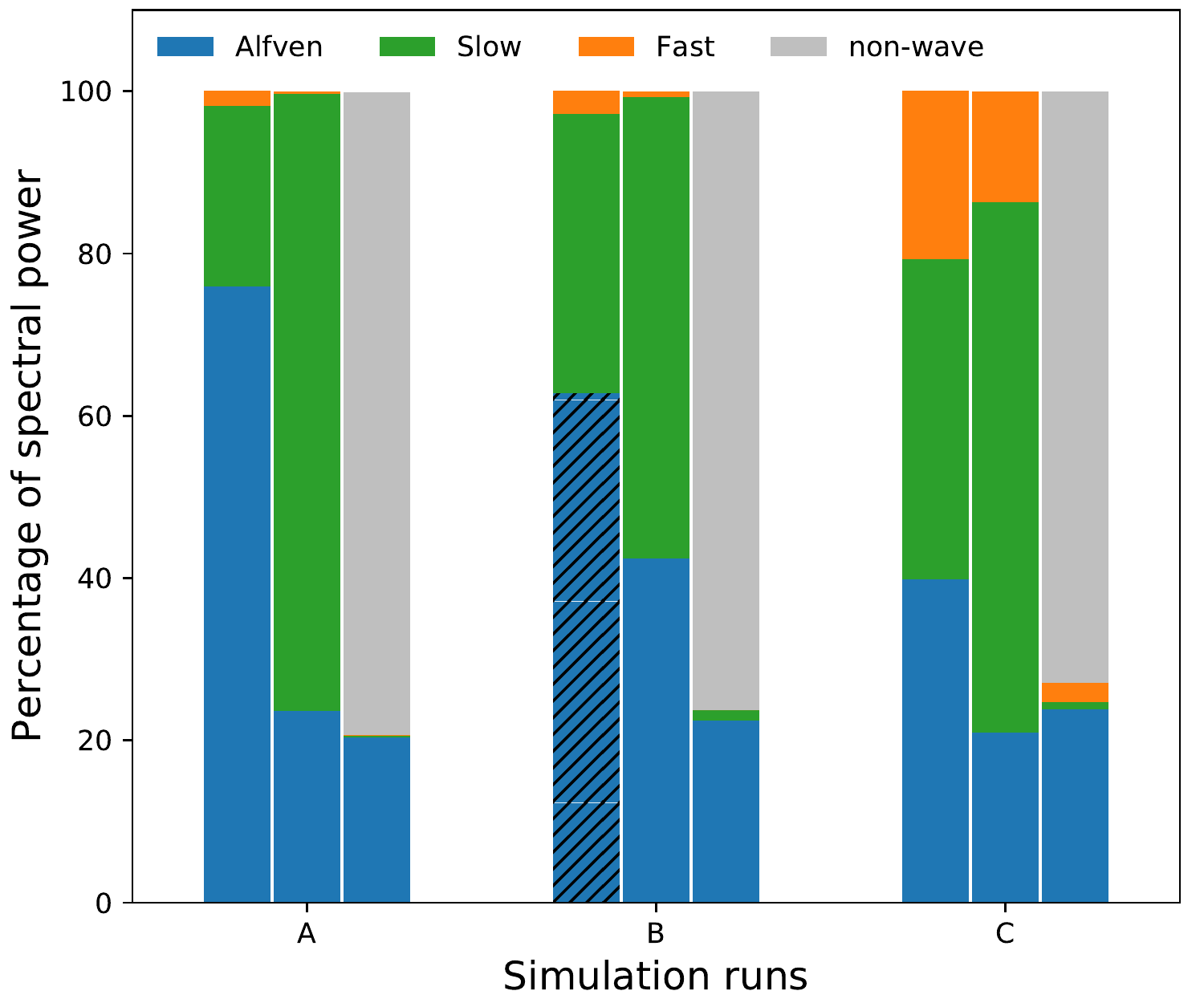}
\caption{Percentage of spectral power derived from the spatial-only 
mode decomposition (left and middle bars) 
and spatio-temporal spectra (right bars)
for runs A, B, and C. The (shaded) left bars use the whole data cubes
where the middle and right bars use data cubes with some excluded
regions (see text for details).
With the spatio-temporal spectra method, the non-wave 
component is dominant and the Fast mode 
fraction is very small (right bars).
}
\label{fig2}
\end{figure}

Second, we quantify the percentage of spectral power in various 
components using runs A, B and C, when the evolution of
each run has reached quasi-steady state 
(after $7-10\tau_A$). Because it is uncertain how to interpret
the power associated with the injection $k-$space, 
we adopt two approaches. The first is to use 
outputs including the injection phase space
and calculate the
percentage of various components using the spatial-only method.
We find that the fractions of A/F/S for run A, B, and C are:
0.760/0.223/0.017 (A), 0.628/0.028/0.344 (B) and 0.400/0.206/0.394 (C), respectively. 
These are represented as the (shaded)
left bars in Figure \ref{fig2}. Both the dominance of Alfv\'en modes
and the increasing fraction of Fast modes with compressible driving
are consistent with the previous studies \citep[e.g.,][]{makwana_properties_2020}. 

The other approach we used
is to exclude certain regions in $k-$space, 
then use both the spatial-only and the 4D FFT methods to 
calculate the power fraction in each component. 
In order to avoid ambiguity, we exclude the following parts from 
our percentage analysis. First,  
because the injection $k$ range 
is dominated by the injection process, we will exclude
the fluctuation power with $|{\bf k}| \leq 2$ in all analysis.
In run B, this is $\sim 70\%$
of the total spectral power in 
the simulation.
In addition, we also exclude fluctuations with $k_\parallel = 0$
and $\omega = 0$
due to the degeneracy of Alfv\'en and Slow modes.
This is $\sim 3\%$ of the total power.
After these modifications to the data outputs, 
in Figure \ref{fig2}, we compare
those derived from the spatial mode decomposition method
(middle bars) and spatio-temporal spectrum method (right bars). 
For the spatial decomposition method, we find the following:
 the A/F/S fractions are: \textcolor{black}{0.237/0.003/0.760 (A),
0.425/0.007/0.568 (B) and 0.210/0.137/0.653 (C)}, respectively.  
The slow mode actually has the largest fraction among three wave
modes using the spatial decomposition method. 
This is different from the previous results by other groups probably 
because we excluded the injection range.  
The Fast mode fraction, being \textcolor{black}{0.3\% and 0.7\%} 
for run A and run B  is negligible when the
driving is incompressible, but becomes a noticeable 
fraction \textcolor{black}{$14\%$} in run C when the driving is highly compressible,
which is expected.

\begin{figure}
\centering
\includegraphics[width=0.45\textwidth]{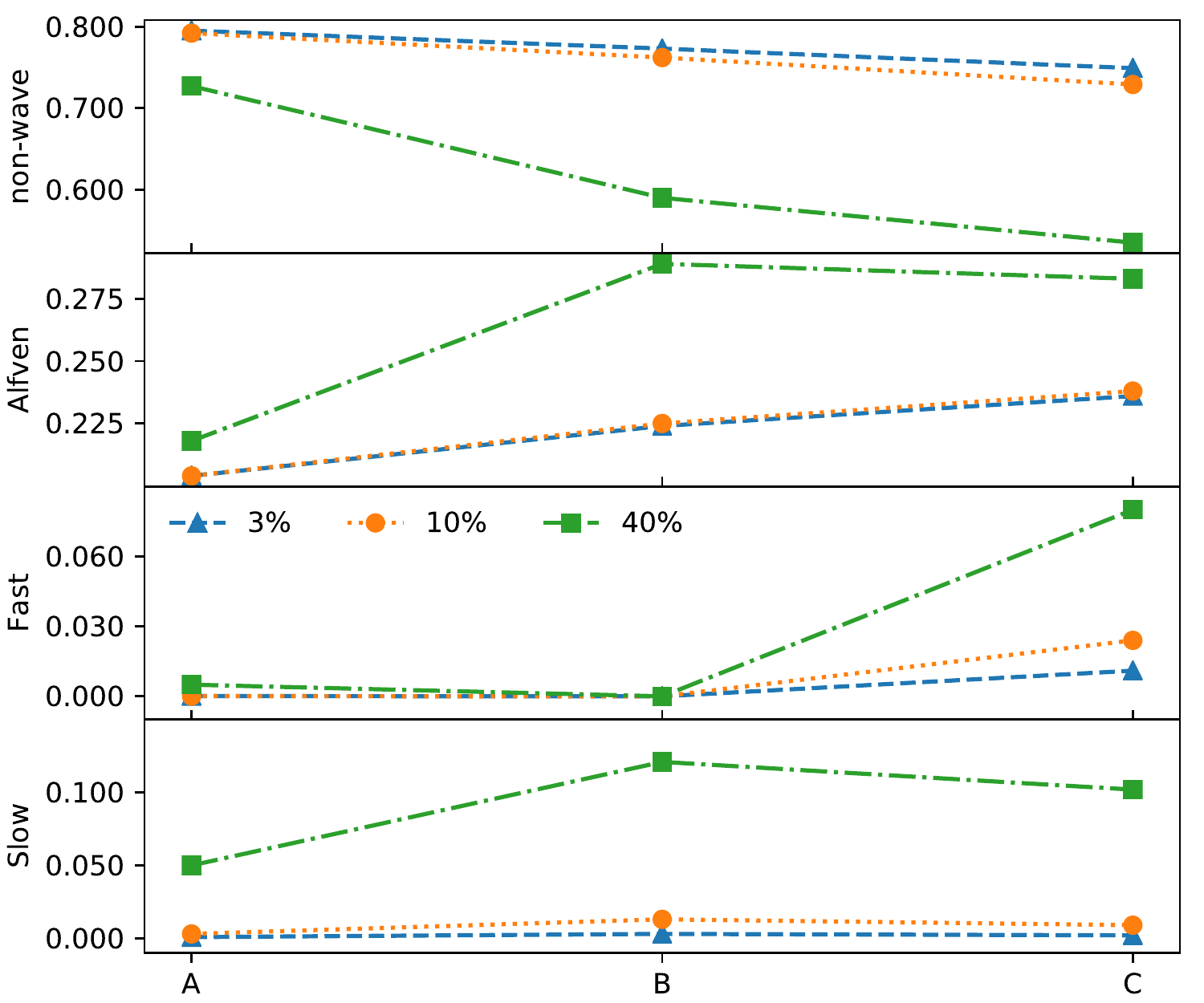}
\caption{Fractions of spectral power for
non-wave, Alfv\'en, Fast and Slow components
(from top to bottom), for run A, B, and C, 
respectively. These
are obtained using the spatio-temporal method only.
}
\label{fig3}
\end{figure}

\begin{figure*}
\centering
\includegraphics[width=0.95\textwidth]{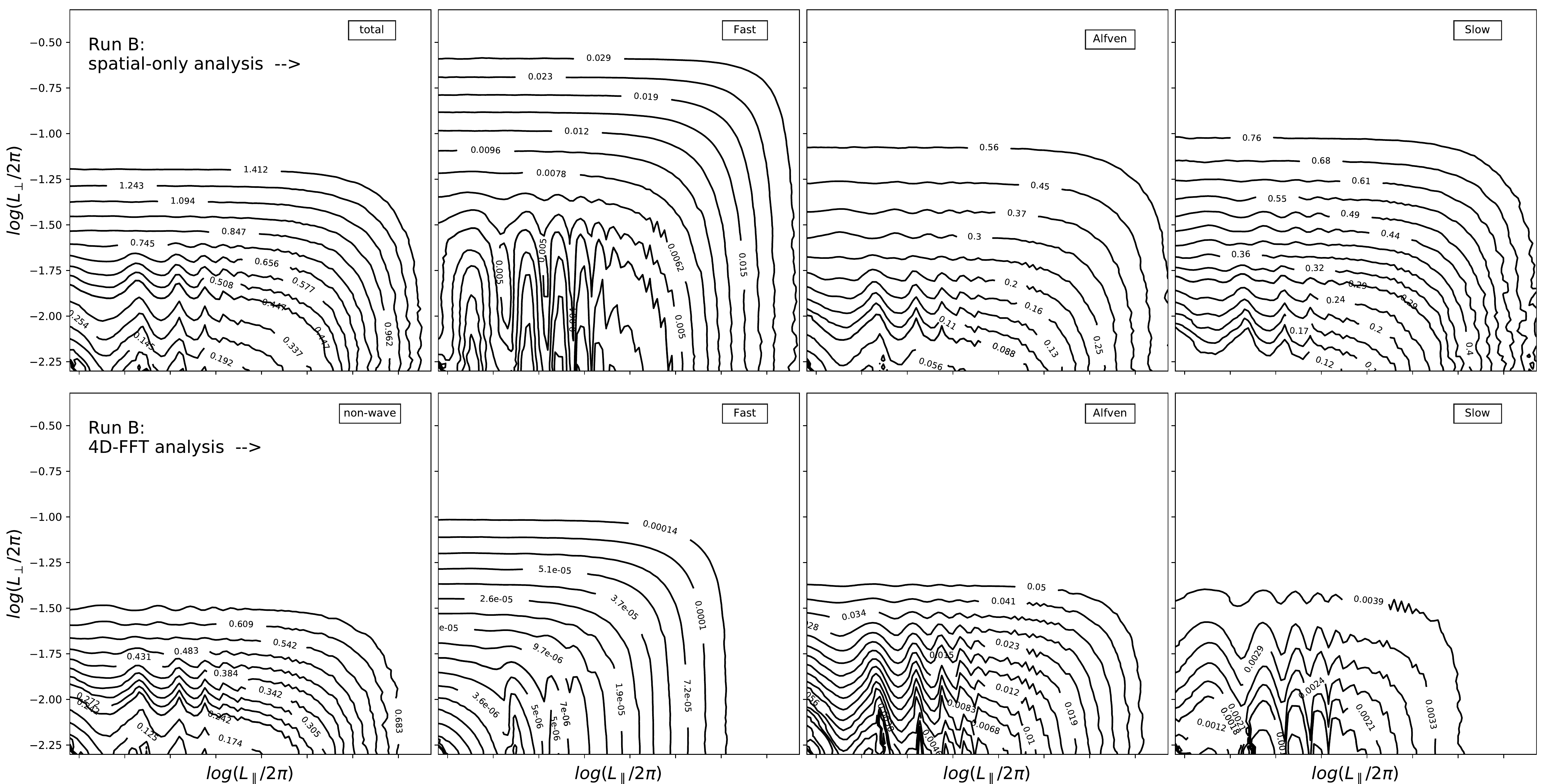}
\caption{Contours of the second-order structure function of velocity 
for run B, separately 
calculated using different components, 
using the spatial decomposition analysis
(top) and the 4D FFT analysis (bottom), respectively.}
\label{fig4}
\end{figure*}

For the 4D FFT approach, we identify various wave branches 
according to the theoretical dispersion 
relations, and we allow 10\% deviation in the theoretical frequencies 
above/below the gray wave surfaces in Figure \ref{fig1}(a) for 
each wave branch. The remaining power which does not fit within
any of the three branches is considered as non-wave. 
Quantitatively, the percentages of these four components 
non-wave/A/F/S are:  
\textcolor{black}{0.792/0.204/$1\times10^{-4}$/0.003 (A),
0.762/0.225/$8\times10^{-7}$/0.013 (B), 
and 
0.729/0.238/0.024/0.009 (C)}, 
respectively.  
They show that the non-wave component is dominant.
With regard to wave modes, the Alfv\'en component 
has the largest fraction. The Fast mode component is 
negligible for both A and B but reaches $\sim 2.4\%$ for C.

One key conclusion demonstrated in Figure \ref{fig2}
is that, while the spectral power can {\it always} be 
decomposed  into one of three ``wave" modes using the
spatial  decomposition method, when taking into account their
frequency behavior, most of these fluctuations do not
follow any dispersion relation. 
Figure \ref{fig2} also shows that the total fraction in waves
using the 4D FFT method 
is less than $\sim 25\%$. 
Run B is designed to have the most Alfv\'enic component
by including magnetic injection.
Indeed, for both methods, the Alfv\'en mode fraction in B
is higher than either A or C.  
This driving also generates
a finite albeit small fraction of slow waves in run B. 
Run C is designed to have the most compressible modes, but
its Fast mode fraction decreases from {$\sim 14\%$} using
the spatial decomposition method to $\sim 2.4\%$ using 
the 4D FFT method.

To test the dependence on the choice of frequency ``width",
we also calculate the percentages of spectral power by 
capturing the power within $\pm 3\%$ and $\pm 40\%$ 
of theoretical frequency, using data cubes with exclusions 
as described above.
These are summarized in Figure \ref{fig3}.
The non-wave/A/F/S components are:
for $\pm 40\%$, 
0.727/0.218/0.005/0.050 (A),
0.590/0.289/$7\times10^{-6}$/0.121 (B), 
and 
0.535/0.283 /0.080/0.102 (C);
for $\pm 3\%$,
0.795/0.204/$1\times10^{-4}$/$8\times10^{-4}$ (A),
0.773/0.224/$2\times10^{-7}$/0.003 (B), 
and 
0.749/0.236/0.011/0.002 (C), respectively.
It can be seen that the overall trends do not 
change as we vary from $\pm 3\%$ to $\pm 40\%$.

Third, we examine the second-order structure 
function of various components
to further quantify their properties. 
We calculate 
the structure functions of individual components in run B
obtained using the two mode decomposition methods
and use $3\%$ frequency width.

Figure \ref{fig4} 
shows the contours of second-order
structure function from run  B using the spatial-only
mode decomposition (top) and the 4D FFT method (bottom). 
For the top row, 
we split the Fourier spectral 
power of the total velocity into Fast, Alfv\'en, and Slow modes. 
Then we make the inverse Fourier transform and calculate the second-order 
structure function of these three modes
in real space. 
Fast modes trend to be isotropic, 
while Alfv\'en and Slow modes are elongated, consistent with previous studies \citep{goldreich_toward_1995,cho_compressible_2002,cho_compressible_2003}. 
To avoid double-counting, we exclude the fluctuations with $k_\parallel =0$ 
(where Alfv\'en and Slow modes are degenerate) and 
$k_\perp =0$ (where Alfv\'en and Fast modes are degenerated). 
For the bottom panel, 
we first identify various wave branches according to the theoretical dispersion 
relation, allowing \textcolor{black}{3\%} deviation in the theoretical frequencies 
above/below the gray wave surfaces in Figure \ref{fig1}(a) for 
each wave branch.  Then we make 4D inverse Fourier 
transform and calculate the structure functions 
for  the non-wave, Fast, Alfv\'en, and Slow components, 
as shown in the contour plots. 
We see the general trend that  
Fast mode appears more isotropic, 
and the other three
components  (non-wave, Alfv\'en, and Slow) are
more anisotropic.
(The slow component might be too noisy to be accurate.)

\begin{figure}
\centering
\includegraphics[width=0.465\textwidth]{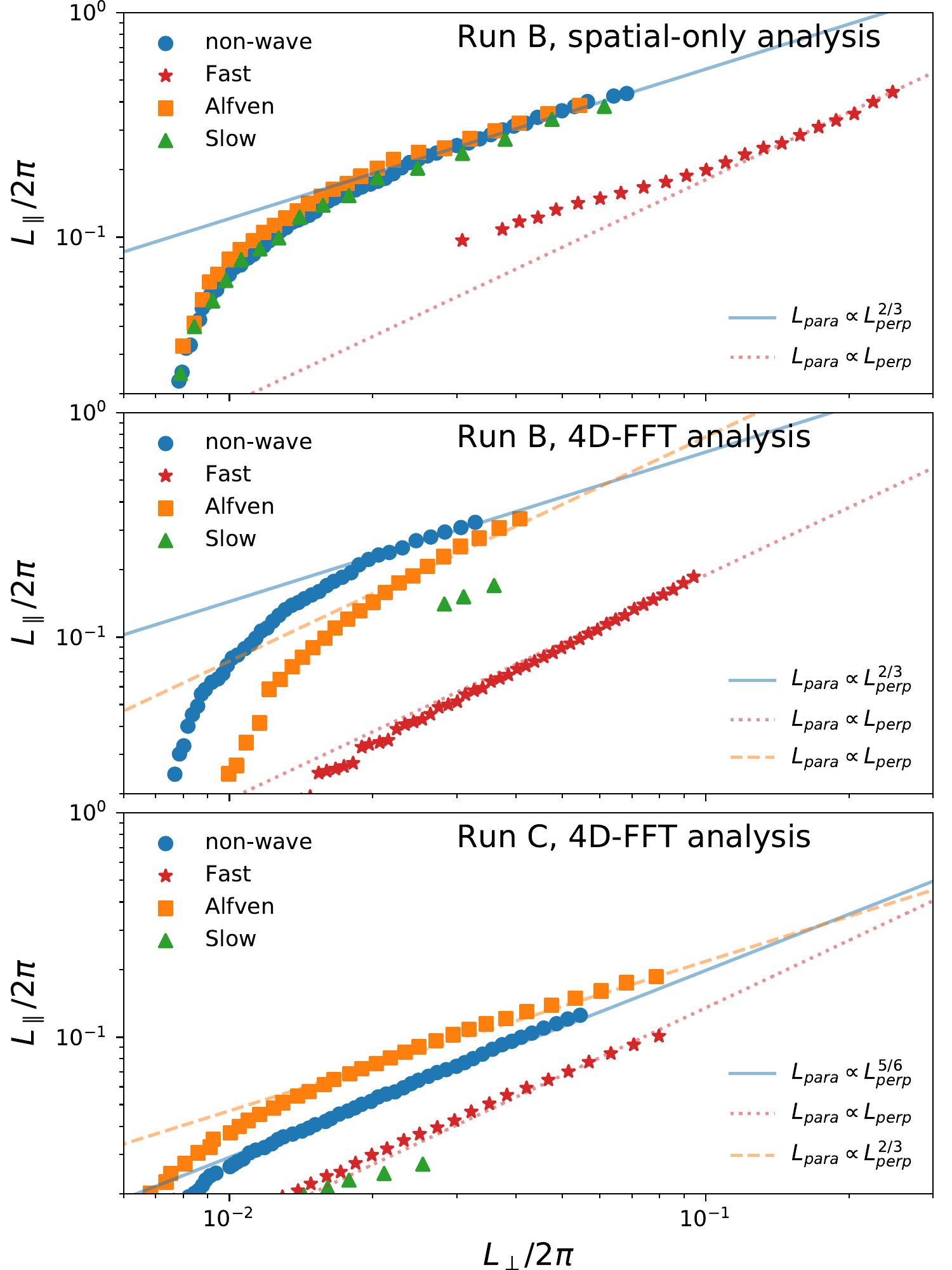}
\caption{Relationships between $L_\perp$ and $L_\parallel$
for different components identified in run B (spatial decomposition only),
run B (4D FFT method), and run C (4D FFT method).}
\label{fig5}
\end{figure}

To further quantify the differences of wave components
derived from the two methods,
in Figure \ref{fig5}, we plot the relationship
$L_\parallel \propto L_\perp^q$ 
of various components identified in runs B and C. 
We summarize the power index $q$  as follows:
Run B spatial-only: total/A/F/S = $\frac{2}{3}$/$\frac{2}{3}$/1/$\frac{2}{3}$;
Run B 4D-FFT: non-wave/A/F/S = $\frac{2}{3}$/1/1/1;
Run C spatial-only, total/A/F/S = $\frac{5}{6}$/$\frac{2}{3}$/$\frac{4}{5}$/$\frac{2}{3}$;
Run C 4D-FFT, Run C 4D-FFT:non-wave/A/F/S = $\frac{5}{6}$/$\frac{2}{3}$/1/1.
(Note that we have added the results
for run C using the spatial-only method.)

Taking Figures \ref{fig4} and \ref{fig5} together,
we can tentatively draw the following conclusions:
1) Using the spatial-only method 
with incompressible driving, 
the relations of $L_\parallel \propto L_\perp^{2/3}$ 
for the total, Alfv\'en and Slow components,  
and $L_\parallel \propto L_\perp$ for the Fast component, are all
the same as those given by previous studies
\citep[e.g.,][]{cho_compressible_2003};
2) Using the spatial-only method but with 
highly compressible driving, the slopes 
for the total ($5/6$) and Fast mode
($4/5$)
components are actually slightly different
from the incompressible case ($2/3$ for total
and $1$ for Fast);
3) Using the 4D FFT method, for A/F/S components,
both run B and 
C give either the same or slightly steeper
slopes from those
obtained via the spatial-only method;
4) Using the 4D FFT method, there are differences
in the slopes for the non-wave and Alfv\'en 
components when comparing run B and C;
5) The Fast component using the 4D FFT method
gives slope of $1$ for both run B and C; 
6) As a consistency check, for both run B
and C, the slopes for
the total ($2/3$, $5/6$) and 
the non-wave component
($2/3$, $5/6$) are the same  because
they dominate the spectral power
using either the spatial-only or the 4D FFT 
method.
These slopes, showing 
slight variations under
different conditions, suggest that
additional theoretical 
and numerical studies 
are needed to address
these differences.

\section{Conclusion and Discussion} \label{sec:discussion}

To understand the nature of MHD turbulence fluctuations 
in the frequency vs. wavenumber
domain in more detail, we have applied both the spatial decomposition
method and the spatio-temporal method to examine the properties 
of MHD turbulence. Particularly, we present results from three
simulations, one (run A) with incompressible velocity driving,
one (run C) with highly compressible velocity driving, and 
one (run B) with incompressible velocity and magnetic field driving. 

After excluding fluctuations associated with driving and
the ($k_\parallel =0,~\omega = 0$) component, we find that:
when taking into account the frequency behavior, 
the majority of the fluctuations cannot fit within any of the 
Alfv\'en, Fast and Slow mode branches. We call them
the ``non-wave" component, which account for about $75-80\%$ of the
total power. 
{ Furthermore, we find that most of the ``non-wave" power is of low frequencies. Similar findings of ultra-low-frequency spectral power were presented in \cite{dmitruk_low-frequency_2007,dmitruk_waves_2009}, i.e., their ``1/f noise''. However, to resolve the ``1/f noise'' in 4D FFT analysis, it may need an even longer time duration and a larger simulation domain than what we used here, which is beyond the scope of this paper.}
{ Observationally, the strong non-wave component is consistent with the results of \cite{bieber_dominant_1996} and subsequent work that showed the solar wind admits frequently an 80\%-20\% decomposition into 2D-slab modes. Theoretically, the nearly incompressible models of MHD predict (for plasma beta $\sim$1 or $<<$ 1) a dominance of 2D over slab fluctuations \citep{zank_nearly_1992,zank_nearly_1993,zank_theory_2017,zank_spectral_2020}. }

For those fluctuations that fit within one of the wave
branches, the Alfv\'en mode dominates. The Fast mode is
essentially negligible in runs with incompressible driving, and becomes
$\sim 2.4\%$ in the run with the highly compressible driving
{ (see, e.g., \cite{zhao_mhd_2021}, for the minority of fast modes in observations)}.

In addition, we find that the second-order structure functions
for different components show differences from those
obtained based on the spatial decomposition method. 

Because the Fast modes in MHD turbulence have been 
postulated to play an important role in understanding
particle transport and energization, our results here should
open up new questions on the existence of these Fast modes
and what quantitative roles they could play. 
Using the spatial decomposition method to identify different
wave modes might be too optimistic in concluding the fraction
of Fast modes (and compressible modes in general). 
Instead, we suggest that the ``non-wave" component needs to be taken into account in the particle transport
and energization processes in MHD turbulence. 
This will be a subject of our future studies.

We acknowledge the support from a NASA/LWS project under award No. 80NSSC20K0377 and 80HQTR20T0027. 
H.L., S.D. and X.F. also acknowledge the support by DOE OFES program and LANL LDRD program.
Useful discussions with X. Li and F. Guo are gratefully acknowledged. 
Simulations were carried out using LANL's Institutional Computing resources.

\bibliography{reference}
\bibliographystyle{aasjournal}

\end{document}